\documentclass[AMA,Times2COL,square,sort&compress]{WileyNJDv5}
\usepackage{multirow}
\usepackage{xcolor}
\usepackage{lineno,hyperref}
\usepackage{gensymb}
\hypersetup{
    colorlinks=true,
    linkcolor=blue,
    bookmarksnumbered,
    citecolor=blue,
    urlcolor=blue,
    bookmarksopen=true
}
\usepackage{titlesec}

\articletype{Research Article}%

\startpage{1}

\begin{document}

\title{Formation of Micrometer-Sized Textured Hexagonal Silicon Crystals via Nanoindentation}
\author[1]{Mouad Bikerouin$^{\bigtriangleup},$}
\author[1]{Anna Marzegalli$^{\bigtriangleup},$}
\author[2]{Davide Spirito}
\author[3]{Gerald J. K. Schaffar}
\author[4]{Corrado Bongiorno}
\author[1]{Fabrizio Rovaris}
\author[4]{Mohamed Zaghloul}
\author[2]{Agnieszka Anna Corley-Wiciak}
\author[1]{Leo Miglio}
\author[3]{Verena Maier-Kiener}
\author[2,5]{Giovanni Capellini*}
\author[4]{Antonio M. Mio}
\author[1]{Emilio Scalise*}

\titlemark{Formation of Micrometer-Sized Textured Hexagonal Silicon Crystals via Nanoindentation}

\address[1]{\orgdiv{Department of Materials Science}, \orgname{University of Milano-Bicocca}, \orgaddress{\state{Via R. Cozzi 55, I-20125, Milano}, \country{Italy}}}

\address[2]{\orgname{IHP-Leibniz-Institut für innovative Mikroelektronik}, \orgaddress{\state{Im Technologiepark 25, 15236 Frankfurt(Oder)}, \country{Germany}}}

\address[3]{\orgdiv{Department of Materials Science}, \orgname{Montanuniversität Leoben}, \orgaddress{\state{Roseggerstrasse 12, 8700 Leoben}, \country{Austria}}}

\address[4]{\orgdiv{Institute for Microelectronics and Microsystems (IMM)}, \orgname{Consiglio Nazionale delle Ricerche (CNR)}, \orgaddress{\state{Strada VIII N. 5, 95121, Catania}, \country{Italy}}}

\address[5]{\orgdiv{Department of Sciences}, \orgname{Università Roma Tre}, \orgaddress{\state{V.le G. Marconi 446 00146}, \country{Italy}}}

\corres{emilio.scalise@unimib.it (Emilio Scalise); \\ capellini@ihp-microelectronics.com (Giovanni Capellini)}



\abstract[Abstract]{We present a comprehensive study on the formation of micrometer-sized, textured hexagonal diamond silicon (hd-Si) crystals via nanoindentation followed by annealing. Utilizing advanced characterization techniques such as polarized Raman spectroscopy, high-resolution transmission electron microscopy, and electron energy-loss spectroscopy, we demonstrate the successful transformation of silicon into high-quality hd-Si. The experimental results are further supported by first-principles calculations and molecular dynamics simulations. Notably, the hd-Si phase consists of nanometer-sized grains with slight misorientations, organized into large micrometer-scale textured domains. These findings underscore the potential of nanoindentation as a precise and versatile tool for inducing pressure-driven phase transformations, particularly for the stabilization of hexagonal silicon. The textured nature of hd-Si also presents a unique opportunity to tailor its optical properties, opening new avenues for its application in semiconductor and optoelectronic devices.}



\maketitle

\renewcommand\thefootnote{}
\footnotetext{$^{\bigtriangleup}$ These authors contributed equally to this work.}

\begin{figure*}[!t]
\centerline{\includegraphics[width=0.8\textwidth]{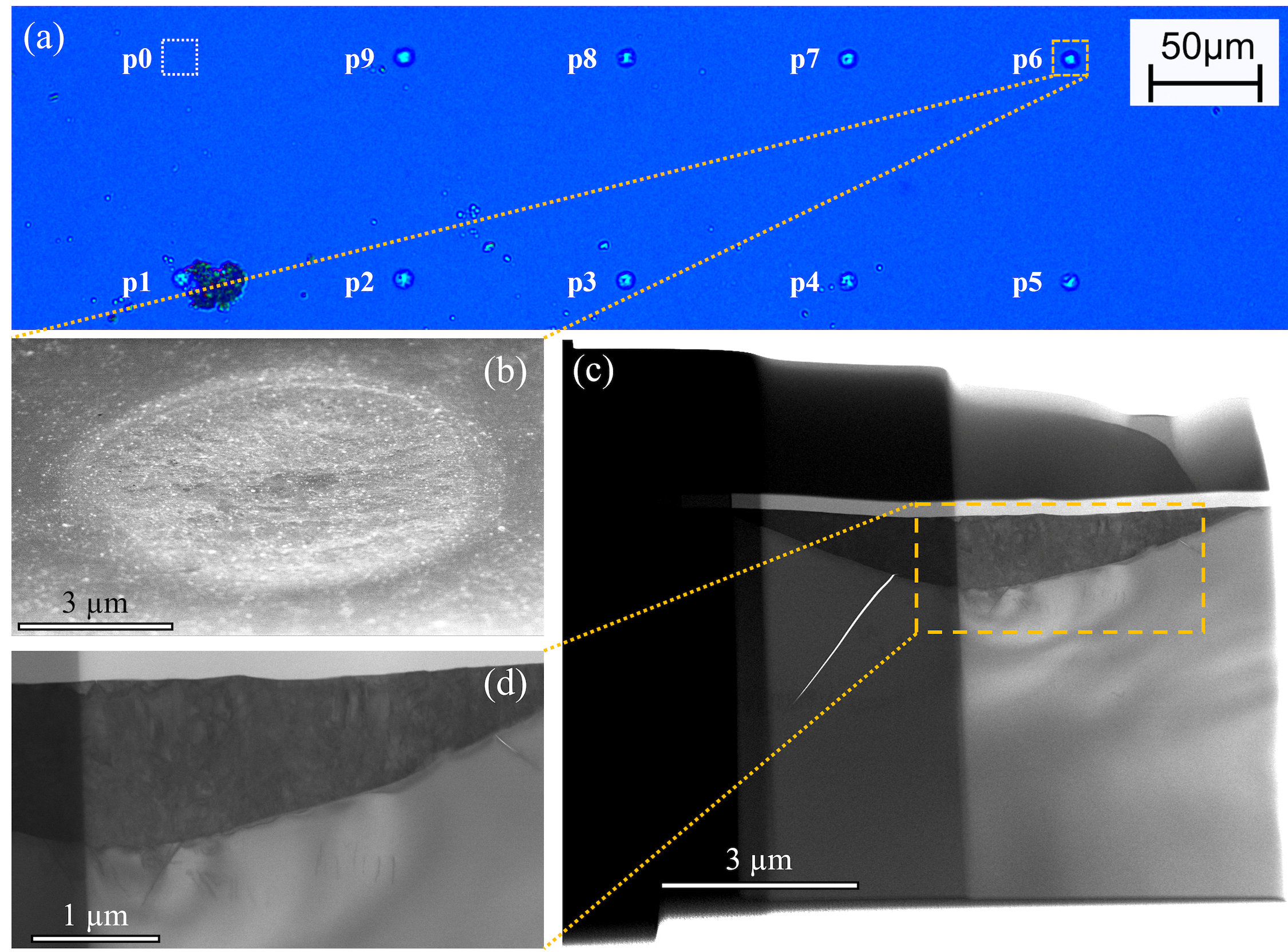}}
\caption{Characterization of silicon post-nanoindentation using a 20 µm spherical tip: (a) Optical micrograph of a series of silicon pits (p1 to p9) formed by nanoindentation, with p0 represents the unindented, pristine dc-Si phase. (b) Scanning electron microscopy (SEM) image of pit p6, highlighted by the yellow dashed box in (a), showing its surface topography. (c) Low-magnification scanning transmission electron microscopy (STEM) bright-field image of the indented region at pit p6, illustrating the protective carbon layer deposited in situ via electron beam and ion beam-induced deposition (EBID and IBID, respectively) and the underlying silicon phases. (d) High-magnification STEM image of the localized region indicated in (c),  revealing further details of the microstructural features within the transformed silicon phase.\label{Fig-Pits}}
\end{figure*}

\renewcommand\thefootnote{\fnsymbol{footnote}}
\setcounter{footnote}{1}
\section{Introduction}\label{sec1}
Silicon (Si), one of the most abundant and globally distributed elements, is the foundational material for complementary metal oxide semiconductor (CMOS) technology, which powers Si-based microelectronics. Within this technology, Si is primarily utilized in its diamond cubic (dc) lattice form, which can be strained under certain conditions. Nevertheless, other Si allotropes offer intriguing potential for addressing the ever-growing demand for enhanced efficiency in data processing and transmission. Many of these Si polymorphs possess unique properties, positioning them as potential low-cost and monolithically integrable solutions. For instance, some Si crystal structures are promising for on-chip energy harvesting devices due to the low thermal conductivity and high thermoelectric efficiency.\cite{du2021thermal, shao2022thermoelectric, liu2021thermoelectric, zhang2018thermoelectric} Other silicon polymorphs also demonstrate high carrier mobility\cite{guo2015new, malone2008ab2} and superconductivity,\cite{malone2008ab1} which could pave the way for innovative electronic and hybrid devices.

Furthermore, some polymorphs have been predicted to exhibit a direct band gap, enhancing light emission and detection capabilities for use in on-chip photonic circuits.\cite{he2016direct, tang2015stable, lee2016ab, lee2014computational, wang2014direct, xiang2013towards, botti2012low, cohen2011wave, saleev2017ab} Among these, hexagonal diamond (hd or Si-IV) stands out as particularly appealing and its realization has been demonstrated in nanowires\cite{he2019direct, tang2017natural} and through advanced deposition techniques.\cite{ahn2021optical} Notably, recent studies have showcased efficient light emission from hexagonal SiGe alloy nanowires, wherein the 2H-Si$_{1-x}$Ge$_{x}$ exhibits a direct band gap that can be continuously tuned over a broad range via composition engineering.\cite{fadaly2020direct,peeters2024direct} The emission yield of these alloys is comparable to that of direct-bandgap III-V semiconductors, highlighting their potential for optoelectronic applications.\cite{wang2021electronic, bao2021toward, fan2020p63, fan2018theoretical} Nevertheless, synthesizing such hexagonal SiGe alloys has proven challenging, as it often requires the use of GaAs core nanowires as a growth template. This limitation underscores the necessity for novel synthesis methods that can directly grow high-quality, direct-bandgap SiGe alloys on silicon substrates.

In principle, hd-Si could serve as a template for the epitaxial growth of hexagonal SiGe alloys, given its pronounced thermal stability, which allows it to maintain structural integrity at temperatures exceeding 700 °C.\cite{wong2019thermal} This makes hd-Si ideal for epitaxy and post-processing in device fabrication. However, synthesizing uniform, high-quality hd-Si crystals remains a significant challenge.
One promising method to access metastable silicon phases, including hd-Si, is nanoindentation, a high-pressure technique that has been effective in exploring the complex phase landscape of silicon, enabling the formation of several metastable phases not typically observed at ambient pressure. Under mechanical pressure, diamond cubic silicon (dc-Si) transitions into a metallic $\beta$-Sn phase at elevated pressures.\cite{gupta1980static, hu1984phases} Upon pressure release, various metastable phases can emerge, depending on the specific experimental conditions. These include the body-centered cubic phase bc8 (Si–III), rhombohedral r8 (Si-XII), two simple tetragonal phases, Si–VIII and Si–IX, as well as Si-XIII, which has an as-yet unidentified crystal structure.\cite{zhao2012tetragonal,mujica2015low,wong2019thermal} Additionally, these metastable phases can undergo further transformations upon annealing.\cite{ge2004thermal, mannepalli2019situ, ruffell2007annealing, ruffell2009annealing, haberl2015thermal}

\begin{figure*}[!t]
\centerline{\includegraphics[width=\textwidth]{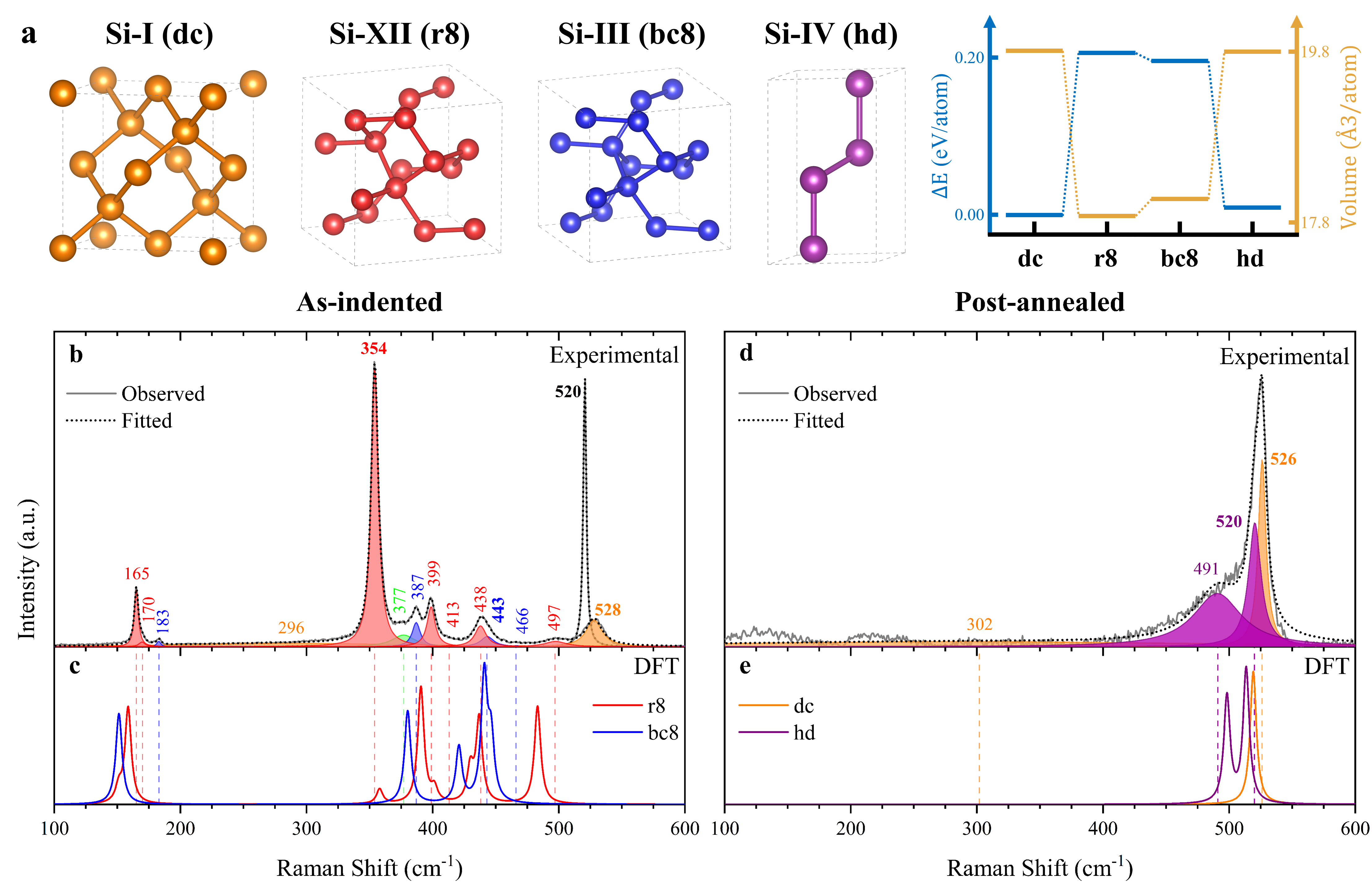}}
\caption{Structural and Raman Spectroscopic Analysis of Silicon Phases Under Nanoindentation and Post-Annealing. (a) Schematic representations of the silicon polymorphs observed in this study: Si-I (dc), Si-XII (r8), Si-III (bc8), and Si-IV (hd). The right plot illustrates the energy difference relative to the dc phase ($\Delta$E in eV/atom) and the equilibrium volume (V in Å³/atom) at zero pressure for each polymorph, emphasizing their relative stabilities and structural similarities. Fitted non-polarized experimental Raman spectra of 20 µm-tip indented silicon: (b) as-indented and (d) post-annealed. The fitting procedure used peak position, intensity, and width as free-fitting parameters. As a reference for the fitting, the frequencies of Raman-active modes for different metastable phases, as reported in previous experimental studies, were employed. Peaks corresponding to various metastable phases are represented by Lorentzian functions: dc (orange), hd (purple), r8 (red), and bc8 (blue). A distinct Raman peak at 377 cm$^{-1}$, indicative of a mixture of r8 and bc8 silicon phases, was identified. For comparison, the pristine Si spectrum is included in the graph. DFT computed Raman spectra are shown for (b) r8 and bc8, and (d) dc and hd. The theoretical spectra were generated using Lorentzian curves with a full width at half-maximum (FWHM) of 3 cm$^{-1}$ to approximately reproduce the experimental FWHM of pristine Si. Experimentally extracted peak positions corresponding to various metastable phases are indicated with dashed lines.\label{Fig-NP-Raman}}
\end{figure*}

By combining nanoindentation with subsequent annealing, it becomes possible to induce phase transformations from the stable dc phase into hexagonal diamond silicon,\cite{wong2019thermal, weill1989characterisation} offering a controlled and scalable approach to hd-Si synthesis. However, the rational design of pressure-induced phase transformations aimed at the specific synthesis of the hd-Si phase remains largely unexplored. A deeper understanding of the mechanisms governing these phase transformations, along with the ability to precisely control these processes, is essential to fully harness the potential of hd-Si for industrial applications. Successfully integrating hd-Si into silicon-based devices using existing fabrication technologies would represent a substantial technological leap, with far-reaching applications in electronics and photonics.

In this paper, we demonstrate the successful realization of high-quality, micrometer-sized, uniform, and textured hexagonal silicon through nanoindentation and subsequent annealing. This achievement is comprehensively demonstrated by combining state-of-the-art characterization techniques, including polarized Raman spectroscopy, high-resolution transmission electron microscopy (TEM), and electron energy-loss spectroscopy (EELS). These experimental results are further supported by first-principles calculations and molecular dynamics (MD) simulations exploiting machine learning-based interatomic potentials. These combined approaches provide a comprehensive understanding of the mechanisms underlying hexagonal silicon formation, opening the door to its potential integration in next-generation semiconductor devices.

\begin{table*}[!t]
\centering
\caption{Comparison of experimental and theoretical Raman frequencies (in cm$^{-1}$) for various investigated silicon phases. Phonon modes have been assigned based on theoretical calculations. Previously reported experimental values are also included for reference. The standard deviation of the frequencies, derived from the analysis of multiple pits, represents the associated errors.\label{table1}}
\begin{tabular*}{\textwidth}{@{\extracolsep{\fill}} cccccccccc @{}}
\cmidrule{1-10}
\textbf{Phase} & \textbf{Raman} & \multicolumn{2}{c}{\textbf{This work}} & \multicolumn{6}{c}{\textbf{Literature}} \\
\cmidrule{3-4}\cmidrule{5-10}
& \textbf{Mode} & \textbf{Exp.} & \textbf{Theo.} & \textbf{Exp.\cite{johnson2011temperature}} & \textbf{Exp.\cite{mannepalli2019situ}} & \textbf{Exp.\cite{smillie2020exotic}} & \textbf{Exp.\cite{wong2019formation}} & \textbf{Exp.\cite{ahn2021optical}} & \textbf{Exp.\cite{hauge2015hexagonal}} \\
\cmidrule{1-10}
dc & T$_{2g}$ & 527.0 ± 3.2 & 519.08 & 520.3 & 516-524 & 522-539 & - & - & 521 \\
\cmidrule{1-10}
\multirow{3}{*}{hd} & E$_{2g}$ & 487.2 ± 7.5 & 498.27 & - & - & - & - & 491.7-494.2 & 496 \\
& E$_{1g}$ & 518.4 ± 6.2 & 513.38 & - & - & - & - & 506.9-513.8 & 514 \\
& A$_{1g}$ & 518.4 ± 6.2 & 513.71 & - & - & - & - & 516.9-519.8 & 514 \\
\cmidrule{1-10}
\multirow{8}{*}{r8} & A$_g$ & 165.2 ± 0.2 & 151.11 & 164.8 & 165.9 ± 1.1 & 162-164 & 165 & - & - \\
& E$_g$ & 170.9 ± 0.9 & 158.53 & 170.0 & - & 169 & 170 & - & - \\
& A$_g$ & 354.5 ± 0.5 & 357.93 & 351.9 & 353.9 ± 0.9 & 357 & 352 & - & - \\
& E$_g$ & 375.6 ± 1.1 & 390.64 & 373.3 & 378 ± 0.9 & - & 373 & - & - \\
& A$_g$ & 399.6 ± 0.5 & 401.34 & 397.1 & 398.5 ± 1.5 & 401 & 397 & - & - \\
& A$_g$ & 415.8 ± 3.0 & 429.79 & 412 & - & 415 & 413 & - & - \\
& E$_g$ & 436.9 ± 2.4 & 436.86 & - & - & - & 433 & - & - \\
& E$_g$ & 495.1 ± 3.7 & 483.16 & - & 495.6 & 490 & 495 & - & - \\
\cmidrule{1-10}
\multirow{5}{*}{bc8} & T$_g$ & 182.8 ± 0.4 & 151.24 & 182.4 & 184 ± 1.2 & - & - & - & - \\
& T$_g$ & 375.6 ± 1.1 & 380.19 & 373.3 & 378 ± 0.9 & - & - & - & - \\
& A$_g$ & 386.6 ± 0.7 & 420.80 & 384.2 & 386.1 ± 1.2 & 388-393 & - & - & - \\
& T$_g$ & 443.1 ± 3.4 & 440.84 & 437.5 & 442.1 ± 1.1 & 443-447 & - & - & - \\
& E$_g$ & 466.0 ± 4.3 & 446.28 & 463 & - & - & - & - & - \\
\cmidrule{1-10}
\end{tabular*}
\end{table*}

\section{Results and Discussion}\label{sec2}
Nanoindentation experiments were performed on monocrystalline silicon (001) using spherical indenters with varying radii and different sets of maximum loads. Following indentation, the samples were annealed in a furnace at 250 °C. The optical micrograph in \autoref{Fig-Pits}a shows the surface imprints produced by a 20 µm indenter tip under a maximum load of 665 mN. Three samples, each containing 9 indentation pits, were prepared under identical conditions, with the detailed analysis below focusing on one representative sample. A magnified view of a single pit, acquired via scanning electron microscopy (SEM), is shown in \autoref{Fig-Pits}b. Additionally, \autoref{Fig-Pits}c,d display scanning transmission electron microscopy (STEM) images of a cross-sectional cut from the same pit, revealing microstructural details of the transformed phases. The adequate choice of the indentation parameters allowed to obtain a uniform region of phase transformation, with a clear absence of cracks at the surface. Moreover, the corresponding load-displacement curves show no evidence of "pop-in" events during loading, suggesting the absence of crystalline defect nucleation and propagation. Out of the 27 total pits analyzed across three samples, 70\% (8 out of 9 pits in two samples, and 5 out of 9 in the third) exhibited these desirable characteristics. In contrast, the remaining pits displayed signs of mechanical instability, with "pop-in" events observed in their loading curves. As further elucidated by the Raman spectroscopy results, the variation in pit morphology and mechanical response is directly linked to differences in the pressure-induced phase transformation behavior. In the majority of pits, silicon reached the critical pressure for phase transformation, resulting in a uniform, micrometer-sized transformed region. However, in the pits that exhibited "pop-in" events during loading, extensive plastic relaxation led to a higher density of defects and cracks, less uniform phase transformations, and smaller transformed regions.

\begin{figure*}[!t]
\centering
\centerline{\includegraphics[width=0.8\textwidth]{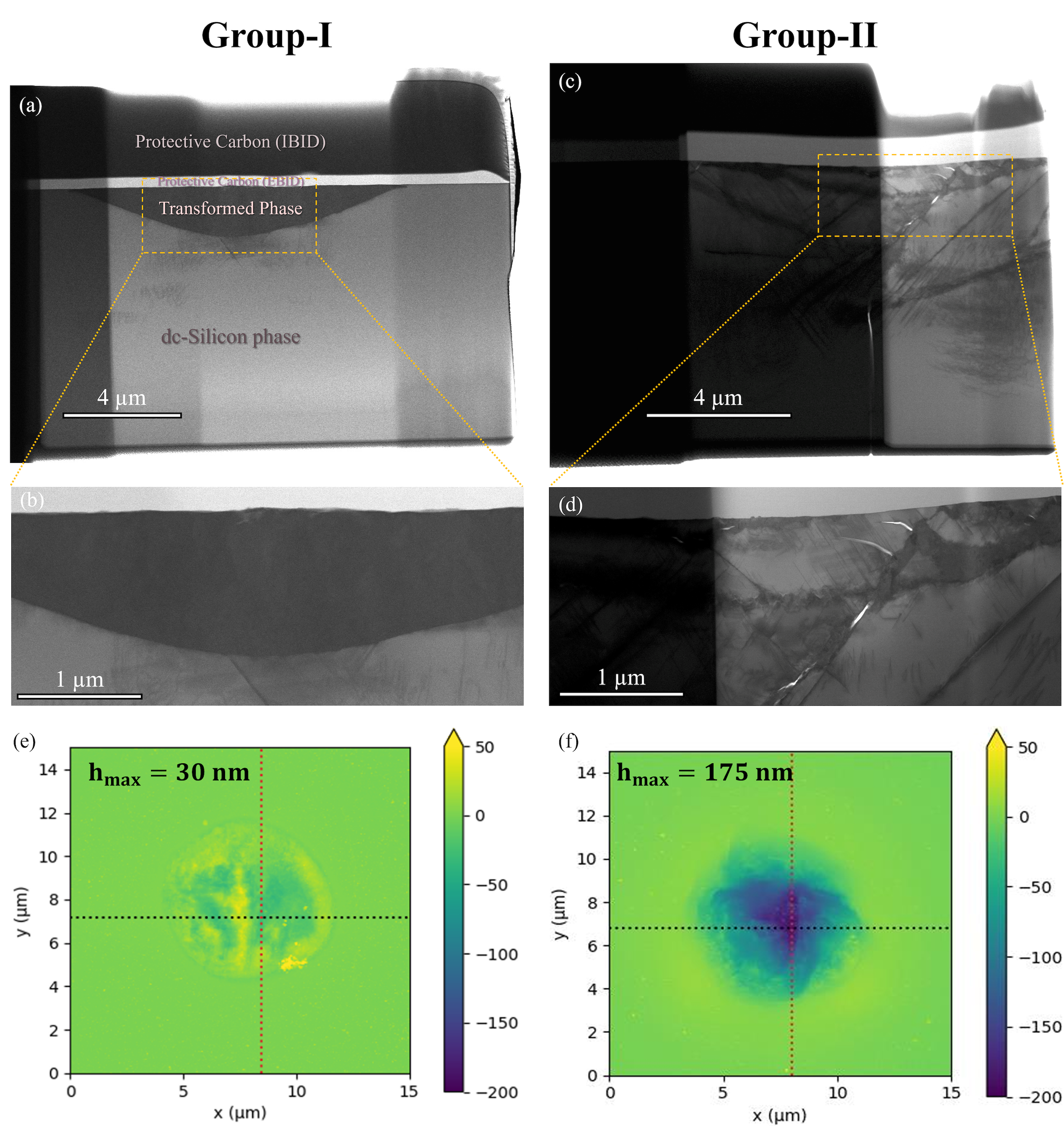}}
\caption{Low-magnification STEM bright-field images of post-annealed silicon indented using a 20 µm tip, showcasing two distinct groups of pits: (a-b) Group-I and (c-d) Group-II. Each sample was protected by in situ deposited carbon layers (EBID and IBID) on the free surface before Focused Ion Beam (FIB) preparation. Images (a) and (c) show overviews of the indented regions, while images (b) and (d) present higher-magnification views of the areas enclosed by yellow dashed rectangles, focusing on the transformed silicon phases. Panels (e) and (f) provide corresponding atomic force microscopy (AFM) topography maps for Group-I and Group-II, respectively, with the maximum height (h$_{\text{max}}$) indicated for each group.\label{Fig-Groups}}
\end{figure*}

To further investigate the structural changes induced by nanoindentation and annealing, we performed a detailed Raman spectroscopy analysis. In \autoref{Fig-NP-Raman}, the non-polarized Raman spectra of both as-indented and post-annealed silicon are presented and compared with theoretical spectra derived from first-principles calculations. Before annealing, the experimental spectrum clearly reveals the presence of the bc8 and r8 silicon phases, with their corresponding Raman peak frequencies listed in \autoref{table1}. The crystal structures of these phases, along with their energy differences relative to the dc phase and their equilibrium volumes, calculated using density functional theory (DFT), are shown in panel (a). As can be revealed from \autoref{table1}, our experimental results are in good agreement with previously reported values in the literature. However, it is worth noting that Raman-active modes associated with the bc8 and r8 phases often display discrepancies across experimental reports. These variations can largely be attributed to the different pressure-inducing techniques employed, such as nanoindentation versus conventional compression experiments.\cite{kobliska1972raman, hanfland1990raman, olijnyk1999effect, kailer1999raman, ruffell2009annealing, haberl2015thermal, wong2019thermal, mannepalli2019situ, smillie2020exotic} These techniques present several challenges, including limited experimental resolution, phase mixtures with overlapping features, and broadened peaks, which complicate the precise identification and assignment of Raman modes. Despite these challenges, our Raman spectra exhibit exceptional resolution compared to previous studies, capturing all Raman-active phonon modes. This enhanced resolution enables clear identification of the r8 and bc8 phases. \autoref{table1} includes the theoretical zone-center Raman-active modes for the r8 and bc8 phases. By focusing on the principal Raman peak frequencies most commonly reported in experiments for each phase, specifically 443.1 ± 3.4 cm$^{-1}$ (T$_{g}$) for bc8 and 354.5 ± 0.5 cm$^{-1}$ (A$_{g}$) for r8, we observe strong agreement with our DFT computations. These Raman modes are characteristic of the bc8 and r8 phases, clearly distinguish them from other possible phases. It is therefore evident that the as-transformed silicon is predominantly composed of the r8 and bc8 phases, with some residual dc-Si also present. For dc-Si, we computed the Raman-active T$_{2g}$ mode, originating from a triply degenerate LO-TO phonon mode, yielding a 519.08 cm$^{-1}$ value. This result closely matches the experimental value of 520.7 cm$^{-1}$, as measured in the pristine (unindented) region of the silicon sample (p0 in \autoref{Fig-Pits}a). For comparison, the corresponding spectrum is also shown in \autoref{Fig-NP-Raman}b. Notably, the dc-Si peak in the indented pits is shifted, indicating the presence of compressive strain in those regions. The conclusions drawn from the as-indented spectra are further supported by Figure S1, where consistent peak positions were observed across different pits, reinforcing the reliability of our findings.

After annealing, significant changes in the Raman spectra are observed, particularly the disappearance of peaks associated with the bc8 and r8 phases and the emergence of new peaks corresponding to the hd phase, indicating complete structural transformations. Specifically, an asymmetric peak with a lower wave-number shoulder is observed around 520 cm$^{-1}$, which is best fitted by a composite of dc and hd phases. Once again, we find good qualitative agreement between the experimental Raman spectra and our DFT computations. All specific Raman peaks for the hd phase are listed in \autoref{table1}. The non-degenerate A$_{1g}$ mode corresponds to the longitudinal optical (LO) phonon mode, while the doubly degenerate E$_{1g}$ and E$_{2g}$ modes correspond to transverse optical (TO) phonon modes. The Raman-active modes for the hd phase were theoretically determined to be 498.27 cm$^{-1}$ (E$_{2g}$), 513.39 cm$^{-1}$ (E$_{1g}$), and 513.71 cm$^{-1}$ (A$_{1g}$). These theoretical values closely align with previously reported experimental Raman frequencies for hd nanowires: 496 cm$^{-1}$ (E$_{2g}$), 514 cm$^{-1}$ (E$_{1g}$), and 514 cm$^{-1}$ (A$_{1g}$).\cite{hauge2015hexagonal} In contrast, our current experimental data show a down-shift of the E$_{2g}$ mode and an up-shift of the A$_{1g}$ and E$_{1g}$ modes, consistent with the presence of uniaxial strain. Notably, in both the theoretical and experimental data, the E$_{1g}$ and A$_{1g}$ modes occur at nearly identical frequencies and are unresolved. The width of the experimental hd-Si peak at 520 cm$^{-1}$ is significantly narrower compared to previously reported hd-Si spectra,\cite{wong2019thermal, ahn2021optical} indicating high crystal quality of the annealed samples. Additionally, no Raman peaks corresponding to amorphous silicon (a-Si) were observed in either the as-indented or post-annealed conditions. Specifically, the characteristic a-Si peaks at approximately 480 cm$^{-1}$ (TO), 380 cm$^{-1}$ (LO), 300 cm$^{-1}$ (LA), and within the range of 80 to 200 cm$^{-1}$ (TA) were absent.\cite{gerbig2015situ} These conclusions hold consistently across all synthesized pits, as demonstrated in Figure S1, further validating the robustness of our results.

It is important to note that, even after annealing, the dc-Si peak in the Raman spectra remains shifted due to residual compressive strain. The dc peak in the spectra exhibits significantly narrower widths, higher intensities, and a more pronounced frequency shift in the pits that underwent plastic relaxation after indentation, corresponding to the 30\% of pits previously identified with "pop-in" events in their loading curves. Henceforth, these are designated as Group-II pits. In contrast, the remaining 70\% of pits, which did not show "pop-in" events during loading, display broader dc peaks, lower intensities, and smaller frequency shifts. These pits are designated as Group-I (see Figure S2 and Figure S3 for comparative Raman spectra and Figure S4 for the corresponding loading curves). After annealing, Group-II pits exhibit a distinct morphology, characterized by the presence of cracks and extended defects, particularly in the dc regions within the transformed zone, as revealed by STEM images in \autoref{Fig-Groups}. The formation of these cracks impedes the full transformation from the dc phase to the bc8/r8 phases, resulting in smaller transformed regions and highly stressed dc areas. As a consequence, this leads to a reduced volume of the hd phase after annealing. This effect is even more pronounced when a smaller tip (10 µm) is used, with nearly 100\% of the pits exhibiting these typical Group-II characteristics, as shown in Figure S6.

\begin{figure}[!t]
\centerline{\includegraphics[width=\columnwidth]{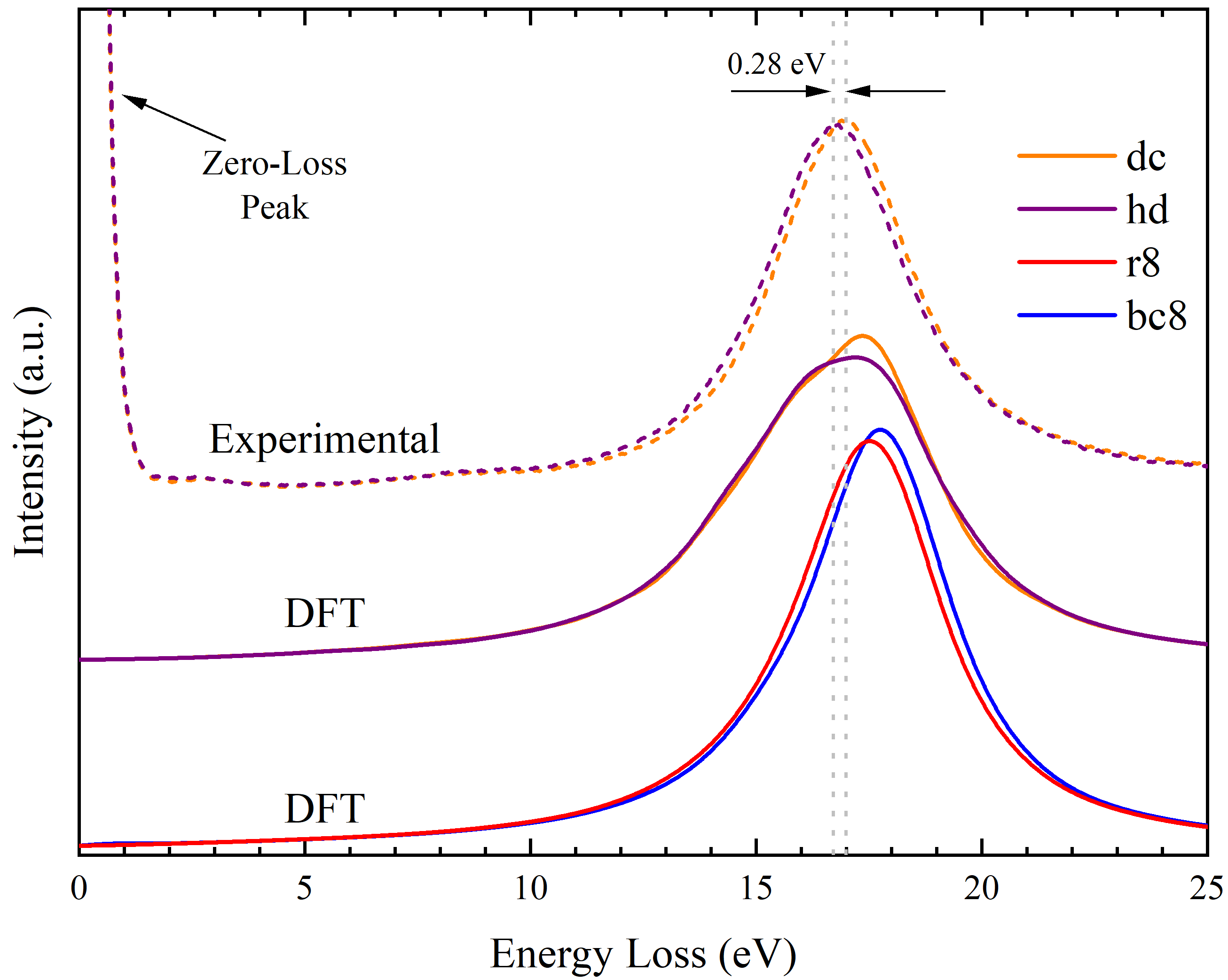}}
\caption{Comparison of experimental and theoretical electron energy loss spectra (EELS) of the low-loss region for post-annealed indented silicon. The experimental spectra (dashed lines) are shown alongside the DFT calculations (solid lines) for the dc (orange), hd (purple), r8 (red), and bc8 (blue) silicon phases.\label{fig:EELS}}
\end{figure}

In contrast, after annealing, Group-I pits, which predominate in silicon samples indented with a 20 µm tip, exhibit large, uniform transformed zones that are free of cracks (see \autoref{Fig-Groups}a,b). Atomic force microscopy (AFM) images also reveal significant differences in the indentation imprint heights between the two groups, as shown in panels (e) and (f), further confirming elastic deformation in Group-I pits and plastic deformation in Group-II pits. Based on the comparison between these two pit families, we propose a distinct approach (detailed indentation parameters are provided in the Experimental Section) aimed at optimizing the statistical occurrence of pits with superior transformed regions. The Raman analysis presented above, along with the forthcoming data, focuses primarily on Group-I pits due to their higher quality and statistical relevance for indentations performed with the 20 µm tip. Complementary data on Group-II pits can be found in the Supplementary Information.

\begin{figure*}[!t]
\centerline{\includegraphics[width=\textwidth]{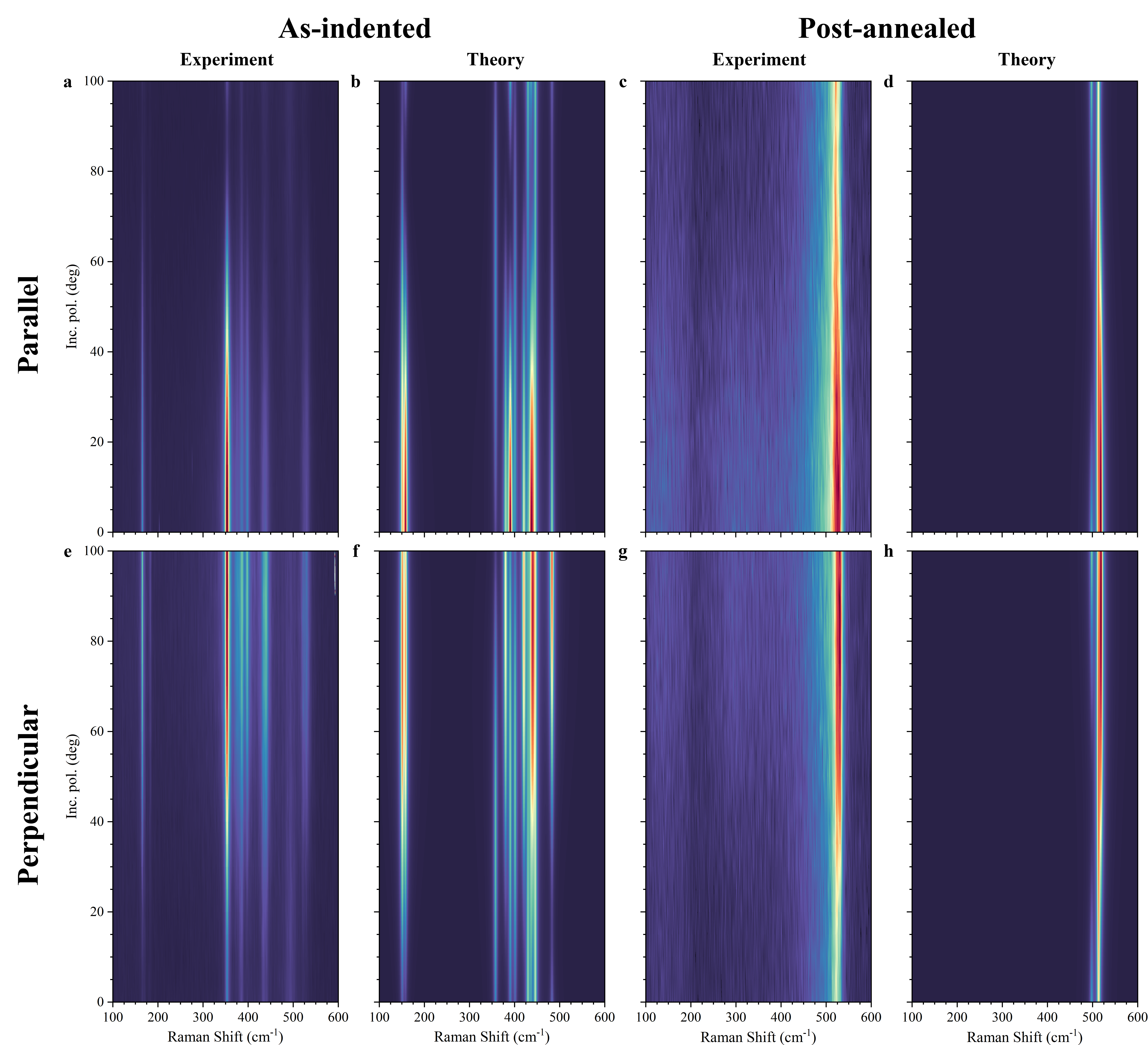}}
\caption{Raman spectra as a function of incidence polarization angle $\beta$ for indented silicon before and after annealing. Panels (a) and (c) show experimental results for as-indented and post-annealed silicon in polarizer-analyzer parallel geometry, respectively. Panels (e) and (g) display the corresponding experimental results in polarizer-analyzer perpendicular geometry. Panels (b), (d), (f), and (h) illustrate the corresponding theoretical predictions. Panels (b) and (f) include the combined theoretical results of the bc8 and r8 silicon phases, while Panels (d) and (h) represent the sum of DFT results for the hd and dc phases. The spectra reveal distinct peaks associated with various silicon phases and highlight the changes induced by annealing.\label{Fig-Ploarized}}
\end{figure*}

\begin{figure*}[!t]
\centering
\centerline{\includegraphics[width=\textwidth]{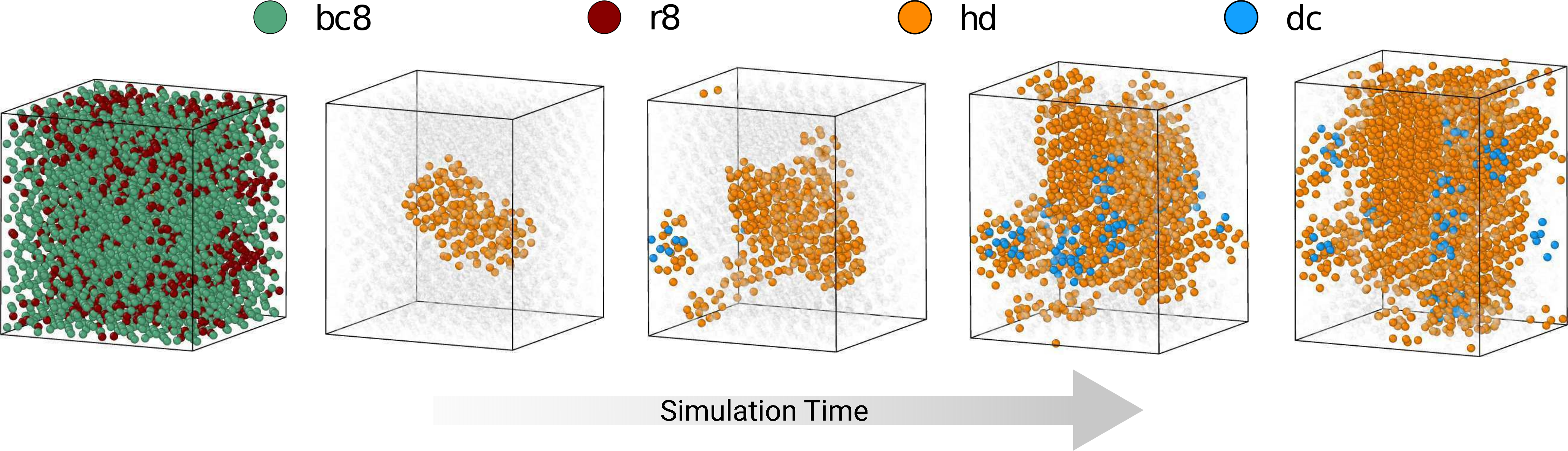}}
\caption{Molecular Dynamics (MD) simulation leveraging a Machine Learning (ML) interatomic potential to illustrate the phase transformation from a mixture of bc8 (green) and r8 (wine) silicon phases to the hd (orange) phase. From the second snapshot onward, bc8/r8 phases are rendered transparent for clarity, allowing for better visualization of the hd phase. The dc phase is represented by blue atoms.\label{fig:nucleation}}
\end{figure*}

To further confirm the complete transformation of the bc8/r8 phases to the hd phase following annealing, the electron energy loss spectrum (EELS) analysis was performed. \autoref{fig:EELS} presents a comparative analysis of experimental and theoretical low-loss EELS spectra for various silicon phases. The key peak observed in the energy loss function corresponds to bulk plasmon excitations in the metastable silicon phases. Experimentally, this peak is observed at an energy of 17.0 eV for the dc phase, whereas the hd phase exhibits a redshift to 16.7 eV, highlighting phase-dependent optical properties. All the salient features observed in the experiment are correctly predicted through DFT computations, with the dc phase showing a peak at 17.36 eV and the hd phase at 17.21 eV. In contrast, the peaks associated with the r8 and bc8 phases are observed at higher energies, 17.52 eV and 17.76 eV, respectively, indicating a blueshift relative to the dc phase. The strong correlation between experimental results and theoretical predictions substantiates both methodologies, affirming the dependability of the techniques employed in this study while providing essential insights for distinguishing various silicon polymorphs. Furthermore, these findings can be exploited in future experimental studies to explore new silicon phases and their associated properties.

In order to conduct an in-depth analysis of the high-pressure metastable silicon phases both before and after annealing, and elucidate their crystallographic orientation, \autoref{Fig-Ploarized} presents the experimental and theoretical polarized Raman spectra as a function of the incidence polarization angle. In general, the visibility and intensity of specific Raman modes can be predicted using polarized Raman selection rules. These rules are dependent on the mode's symmetry and scattering geometry. The Raman intensity of a particular vibrational mode is proportional to the product of incident and scattered photon polarization vectors, as well as the Raman susceptibility tensor. According to group theory, the form of the Raman tensor, which includes the non-zero matrix elements, is determined by the point group and the structure's symmetry. However, the specific values of the matrix elements are determined by the electronic polarizability of the material system under investigation. The Raman tensors are typically presented in principal axis reference systems, but the direction of photon polarization is described in terms of the sample's physical orientation. By examining the visibility and azimuthal dependence of the intensity of a Raman peak with respect to the polarization of the incident and scattered photons, it is possible to determine not only the crystal phase of a material but also its orientation. In our investigations, the analyzer is aligned with the incident polarization either at 0° (parallel polarization) or at 90° (perpendicular polarization). Both experiments and calculations for polarization-resolved Raman scattering were performed in the same back-scattering geometry. The excitation and detection polarization configurations are expected to not affect the Raman shifts, but the relative intensity of the Raman modes will. The theoretical spectra as-indented (post-annealed) are obtained by combining the contributions due to the r8 and bc8 (dc and hd) phases. The exceptional agreement between experimental and theoretical spectra reinforces the conclusions drawn above by the analysis of the non-polarized Raman spectra: r8 and bc8 phases dominate the as-indented transformed phase, while after annealing only dc and hd-Si are detected. Especially for the r8 and bc8 phases, having different Raman modes compatible with other metastable phases, the analysis of the polarized Raman spectra shown in \autoref{Fig-Ploarized} unambiguously evidences the formation of these two phases but excluding other possible metastable phases. Moreover, these two phases show coherence in their crystallographic orientation and their Raman signatures are always coupled, suggesting that the formed r8/bc8-Si is a crystalline mixture of the two phases appearing together,\cite{Zarudi} but not two independent phases as can be formed in a diamond anvil cell (DAC).\cite{crain1994reversible, piltz1995structure, wippermann2016novel, haberl2016pathways}

Additionally, the scattering intensities of the Raman-active modes predicted and measured before annealing (see \autoref{Fig-Ploarized}) are highest for polarization perpendicular to the indentation axis and lowest for parallel polarization, at 0° of incidence polarization angle, while it gets the contrary approaching 90° of incidence polarization angle. Thus, both phases have coherent dependence on the intensity of the Raman peaks with respect to the polarization of the incident and scattered photons. A similar behaviour is observed for the spectra measured after annealing and compared to the simulations for hd and dc phases, suggesting a preferred crystallographic orientation of all metastable phases. This is confirmed in Figures S1, where the same peak frequency, FWHM, and dependence on different polarization configurations were observed in other pits, indicating the common crystalline orientation of all metastable-Si structures formed in each pit, ruling out a polycrystalline nature. Indeed, all these metastable phases follow the selection rules predicted theoretically, asserting effective control and high reproducibility across different pits and samples in the nanoindentation process, which leads to the pressure-induced phase transition, as well as in the subsequent phase transformation upon annealing. Moreover, these results enable polarized Raman spectroscopy to be a valuable tool to easily and accurately distinguish the different metastable phases in Si. Similar conclusions about the presence of bc8, r8, hd, and dc phases, together with their crystallographic orientation, may be extended to the case of the 10 µm tip (see Figures S7-S8), but the transformed zones are considerably smaller (nm scale) because the formation of defects and cracks inhibits the phase transformation over large (µm scale) and uniform area.

\begin{figure*}[!t]
\centering
\centerline{\includegraphics[width=\textwidth]{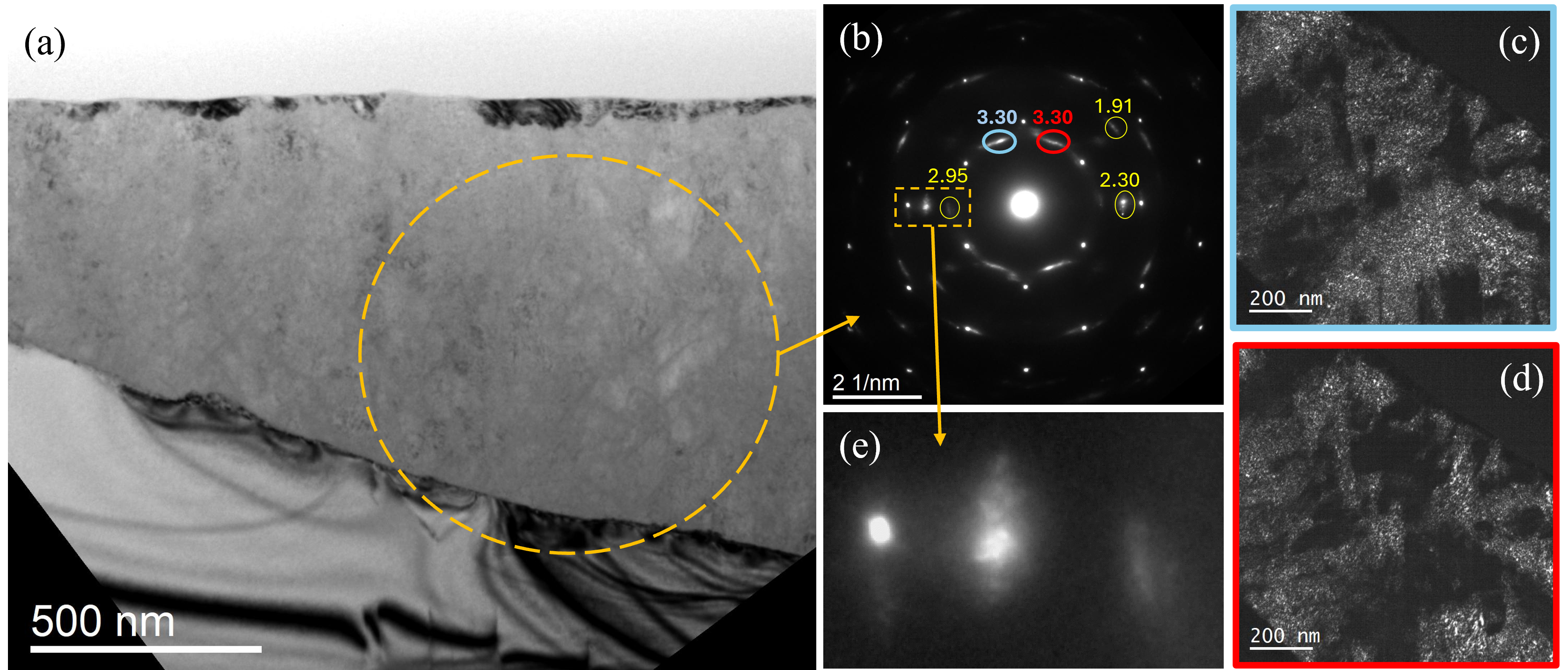}}
\caption{(a) TEM micrograph of the pit in cross-view. (b) Large SAED acquired at the pit center. Typical (hd) 2H-Si reflections are highlighted. Brighter and ordered spots come from the silicon substrate. (c, d) Dark field image acquired using the two arc-shaped spots at 3.30 Å in the same sample area. The images are complementary and cover almost all the sample region, demonstrating the presence of crystal texturing. (e) Magnified view of some electron diffraction reflections, exhibiting a diffused and arc-shaped hd-Si spotty structure.\label{Fig-TEM}}
\end{figure*}

Summarizing the polarized and non-polarized Raman results, we can conclude that the as-indented sample primarily consists of a bc8/r8 composite with high crystallographic coherence, except for some original dc phase with residual compressive stresses. Upon annealed, this bc8/r8 mixture transforms into an hd phase, possibly with some marginal recrystallization of dc-Si. Since the entire bc8/r8-Si phase mixture fully transforms into another phase after annealing, this suggests that r8/bc8-Si is a single hybrid phase rather than two distinct phases. This hypothesis is further supported by molecular dynamics (MD) simulations exploiting a Machine Learning (ML) interatomic potential~\cite{BartokPRX2018,GeActaMat2024}, which were employed to investigate the transition process from the bc8/r8 mixture to the hd phase. Details regarding the simulation setup and the analysis of the simulation output are detailed in the Methods Section. The crystallographic structures of the four phases involved in the simulation are illustrated in \autoref{Fig-NP-Raman}. Selected snapshots from the MD simulation are shown in \autoref{fig:nucleation}. The initial simulation cell, comprising 3456 atoms, was initialized with the bc8 crystal structure and thermalized for 20 ps at 750 K without any applied external pressure. During this thermalization step, a dynamical mixture of the bc8 and r8 phases is naturally formed, yielding a configuration similar to the first snapshot in \autoref{fig:nucleation}. Remarkably, it is not possible to recognise distinct nuclei of the two phases, but a homogeneous mixture of the two phases appears after the thermalization at 750 K. This result suggests that it would be improbable to obtain distinct and large nuclei of bc8 and r8-Si at finite temperature and negligible residual stress, but a hybrid phase resulting from a mixture of the two phases.\cite{rovaris2024unraveling} Note that, a temperature higher than room temperature has been used to speed up the thermalization in the MD simulation, but qualitatively similar results, except for a slight difference in the atomic percentage composition of bc8/r8, are expected for longer simulations using near-room temperatures and larger cells. Subsequently, a phase transition from the bc8/r8 mixture to the hd phase can be identified during the simulation, with the formation of an initial nucleus, as shown in the second snapshots reported in \autoref{fig:nucleation}. Successively, this initial nucleus of the hd phase grows and occupies almost completely the simulation cell, as can be observed in the latest snapshots of \autoref{fig:nucleation}. These snapshots also evidence how some of the starting bc8/r8 phase transforms back toward the most stable dc phase, as evidenced by some small nuclei of dc-Si in the simulation.

To further validate our previous conclusions and gain deeper insight into the microstructural characteristics of the hd-Si phase formed post-annealing, we present the TEM analysis results in \autoref{Fig-TEM}. At first glance, the diffraction contrast image of the pit in cross-sectional view (\autoref{Fig-TEM}a) reveals relatively uniform contrast, with no visible grain boundaries or large domain contours. This uniformity suggests that the domain orientation in the modified area does not follow the random distribution typically seen in polycrystalline materials. Instead, it points to a preferential orientation within large domains, likely influenced by the initial single-crystalline substrate. These findings align well with the polarized Raman results.

\begin{figure*}[!t]
\centering
\centerline{\includegraphics[width=\textwidth]{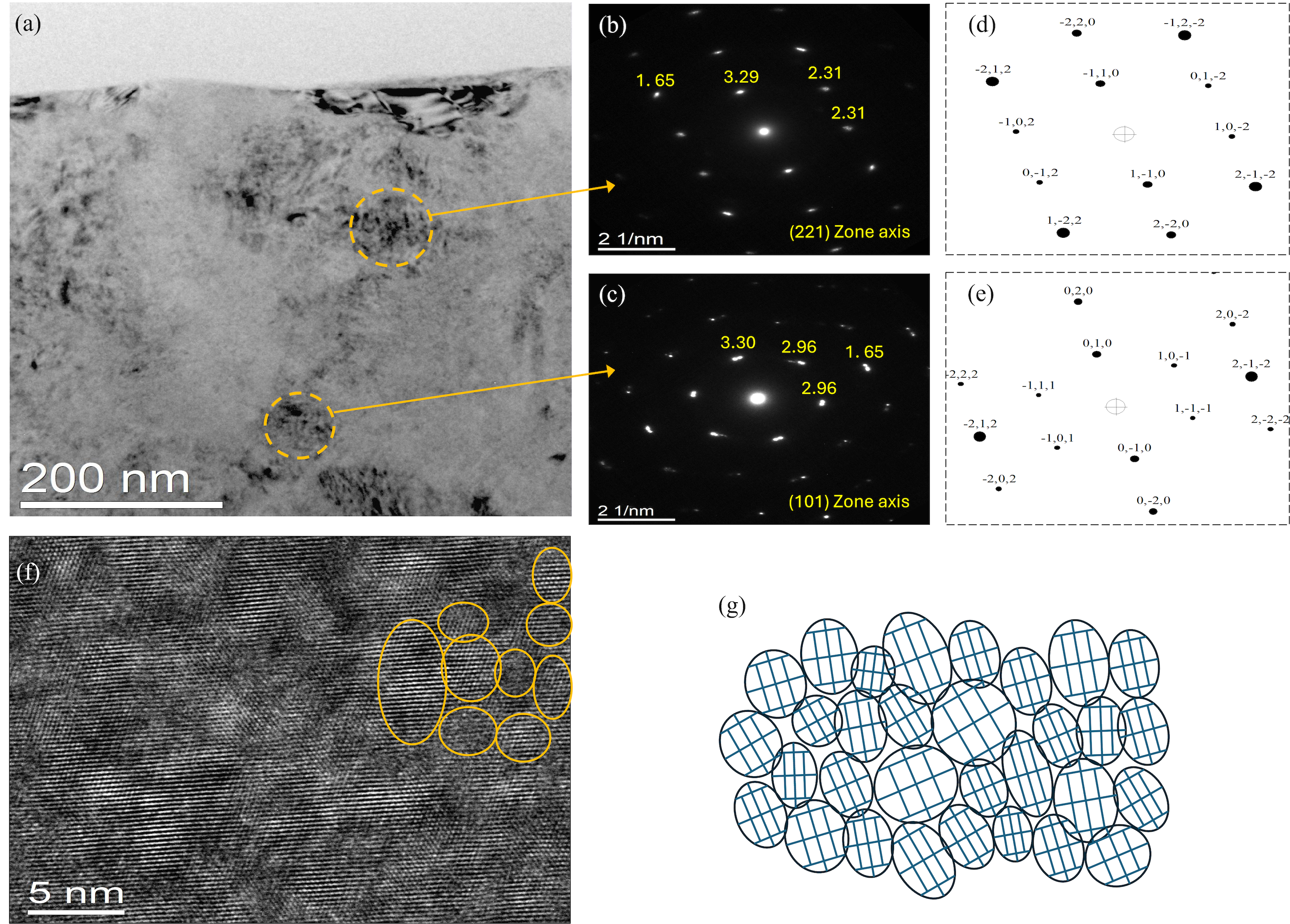}}
\caption{(a) Cross-view TEM micrograph showing large crystal domains inside the pit. (b, c) Diffraction patterns from the domains in the zone axes. (d, e) Diffraction pattern simulations of the hd-Si zone axes. (f) HR-TEM  image of a single domain, composed of nanometre-sized crystals with similar orientation but slightly tilted each other. (g) Schematic description of the crystal texture.\label{Fig-Texture}}
\end{figure*}

The large-area SAED pattern acquired at the pit center, shown in \autoref{Fig-TEM}b, exhibits diffuse, arc-shaped contributions alongside the dc-Si substrate spots. All extracted plane distances from these signals are consistent with the 2H-Si crystal structure but they are incompatible with standard dc-Si. This evidence supports the previous Raman and EELS measurements confirming the presence of hd-Si crystals and further indicates that nearly the entire pit displays diffraction spots corresponding to the hd phase. Meanwhile, the dc phase is confined to small, discontinuous regions approximately 50 nm thick layer at the surface. The discrete and ordered angular distribution of these signals suggests a textured organization of the hd-Si crystals, as confirmed by the dark field images in \autoref{Fig-TEM}c,d. These images, obtained by selecting specific diffraction signals, highlight the areas of the sample contributing to those reflections. The bright regions in these images are complementary, covering almost the entire area, which suggests strong texture in the hd-Si crystal, forming macro-domains with a common orientation. However, the fragmented contrast within each macro-domain is clearly visible, indicating a nanostructured material rather than large monocrystalline domains. This observation is corroborated by the very diffuse spots in the diffraction pattern \autoref{Fig-TEM}e, which are typically associated with nanometer-sized grains. Therefore, macro-domain texturing alone does not provide a complete description of the system.

By tilting the sample and enhancing diffraction contrast using a smaller objective aperture, medium-sized domains within the macro-domains become evident, each showing fringed contrast, as shown in \autoref{Fig-Texture}a. Proper crystal alignment within these domains reveals clear zone-axis diffraction patterns in SAED, as illustrated in \autoref{Fig-Texture}b,c, which correspond well with the hd-Si crystal structure, as confirmed by simulations in \autoref{Fig-Texture}d,e. High-resolution micrographs reveal nanometer-sized grains within these medium-sized domains (\autoref{Fig-Texture}f), most of which share the same crystal orientation but exhibit slight misalignments. These misorientations, typically on the order of a few degrees, are enough to disrupt constructive interference between grains without introducing distinct diffraction spots beyond the average zone-axis.
The combination of small grain size and slight misorientation produces highly diffuse, arc-shaped spots in diffraction, while the cumulative effect of these tilted grains generates a single, clean zone-axis. Many medium-sized domains likely share a common orientation, producing the macro-domain dark field images. The small grain size and misorientation naturally lead to a significant amount of distorted silicon structures, both within the textured nanocrystals and at the boundaries between these nano-grains.

The TEM analysis clearly demonstrates the formation of large, micrometer-sized hexagonal silicon crystals beneath the indentations after annealing. However, it also reveals that the induced hd phase is neither monocrystalline nor polycrystalline. Instead, the hd-Si consists of textured nanocrystals organized into domains, which may exhibit slight misalignment (see \autoref{Fig-Texture}g). The presence of this texturing in both Group-I and Group-II pits, formed using a 20 µm spherical indenter, as well as in smaller 10 µm pits, suggests that this texture is likely an intrinsic feature of the pressure-induced and thermally-driven phase transformation process.

Crystallographic texturing has been observed in several high-pressure silicon phases, including $\beta$-Sn,\cite{jamieson1963crystal} Imma,\cite{mcmahon1993new,mcmahon1994pressure} and sh,\cite{olijnyk1984structural} during pressure-induced deformation. Recently, similar findings have been extended to the bc8 phase, which forms under non-hydrostatic compression using a DAC at 20 GPa.\cite{butler2023origin} In this case, bc8 crystals orient their <110> directions perpendicular to the compression axis but are free to rotate around this direction due to in-plane symmetry inherited from the sh phase. The preferred orientation originates from the highly anisotropic Young’s modulus of the sh phase and is transferred to successive daughter phases via displacive phase transformations. This explains the crystallographic texturing of the hd crystals (the daughter phase) observed in this study, which arises from the transformation pathway involving the bc8/r8 mixture (the parent phases). Indeed, our TEM analysis of the transformed phase (as-indented) confirms the presence of nanometer-sized, slightly misoriented grains, closely resembling those observed in the textured hd-Si after annealing.

\section{Conclusion}\label{sec3}
In conclusion, our study not only provides new insights into the structural evolution of high-pressure silicon phases but demonstrates, for the first time, the successful formation of high-quality, micrometer-sized hexagonal silicon (hd-Si) crystals through nanoindentation. The hd-Si crystals exhibit remarkable uniformity and consistency across multiple indentation pits. The high quality of the hd-Si phase is evidenced by the presence of large and uniform transformed silicon domains, free from cracks and with minimal defects. In rare instances, defects and cracks within certain pits degraded the material’s quality, inhibiting the formation of large hd-Si domains. This thorough comparative analysis of pits formed under varying nanoindentation conditions further provides critical insights into optimizing the indentation parameters necessary to achieve superior material quality.

Notably, our findings further reveal that the hd-Si phase is not a monocrystalline structure, but rather a textured assembly of nanometer-sized grains with slight misorientations, an intrinsic feature of the pressure-induced and thermally driven phase transformation process. This texture introduces opportunities to enhance the properties of hd-Si through the manipulation of anisotropy, strain, quantum confinement, and other novel effects, thereby expanding the possibilities for tailoring its properties to specific applications. Moreover, the textured hd-Si phase presents a promising platform for the epitaxial growth of hexagonal germanium (Ge) and SiGe alloys, which are particularly attractive for optoelectronic applications due to their direct bandgap properties, further broadening the scope of materials that can be integrated into future device technologies. Overall, this work highlights the versatility of nanoindentation as a precise tool for inducing controlled phase transformations in silicon and underscores the potential of hd-Si for future applications in semiconductor technology, optoelectronics, and advanced materials engineering.

\section{Experimental Section}\label{sec4}

\textit{Nanoindentation}: Indentation experiments were performed using a KLA Corporation (Milpitas, CA, USA) G200 System, equipped with spherical diamond tips of 10 µm and 20 µm radii, sourced from Synton-MDP (Nidau, Switzerland). The loading was performed at a constant strain rate of $\dot \epsilon=10^{-3} \; s^{-1}$, following the methodology outlined by Leitner et al. \cite{LEITNER2018, Kiener2023}. The unloading was performed at a constant unloading rate of $\dot P=1 \; mN/s$. This slow unloading was performed to favor crystalline transformation rather than amorphization.\cite{lin2020temperature, domnich2002phase, Schaffar2022, Juliano2004} For indentations performed using the 20 µm radius tip, experiments were carried out up to the system’s maximum applicable force ($P_{max}=665 \; mN$), ensuring sufficient force to trigger the expected phase transformations. In contrast, indentations with the 10 µm radius tip were conducted across three different maximum loads. One sample was subjected to the same maximal load as for the 20 µm tip ($P_{max}=665 \; mN$), while the other two samples were indented with reduced maximum loads of 570 mN and 425 mN, respectively. Despite the reduced force, the smaller tip applied larger local strains compared to the 20 µm indenter due to its geometry, leading to more pronounced mechanical responses. The most significant results obtained using the 10 µm indenter are presented in the supporting material, as the transformed phases generally exhibit lower quality compared to those produced with the 20 µm tip.

\textit{Raman Scattering Experiments}: Raman spectroscopy was performed using a Renishaw inVia system, with an excitation laser wavelength of  532  nm in back-scattering geometry, with the laser incident along the sample crystalline  direction [001] (z) of the substrate. The system is equipped with a polarizer for the excitation laser and an analyzer at the spectrometer's entrance. Additionally, an half-wave plate allows to rotate the polarization of the excitation by an angle $\alpha$. Another half-wave plate allow to set the analyzer direction so that it can be parallel to $\alpha=0$ deg or $\alpha=90$ deg. The sample is aligned so that the direction x'=[110] of the substrate is aligned with the direction $\alpha=0$ deg.

\textit{TEM, SAED, STEM, EELS}: Conventional Transmission Electron Microscopy analyses in parallel beam and Selected Area Electron Diffraction (SAED) were performed by using a JEOL JEM-2010F microscope, equipped with a Schottky field emission gun and operating at 200 keV.
Electron energy loss spectroscopy (EELS)  were performed by using a JEOL ARM200F Cs-corrected microscope, equipped with a cold-field emission gun and operating at 200 keV. A GIF Quantum ER system was used for EELS measurements. EELS spectra were acquired in spectrum imaging (SI) mode.
TEM lamellae were prepared by focused ion beam (FIB) using a Thermofisher Helios 5 UC+ system, operating with 30keV Ga+ and finally polishing at low energy (2keV Ga+) to reduce residual FIB-induced amorphization.
Scanning Electron Microscopy (SEM) and Scanning Transmission Electron Microscopy (STEM) have been performed by using the electron column of the same system.

\textit{MD Simulation}: Molecular Dynamics simulations have been performed with the LAMMPS code~\cite{ThompsonCPC2022} exploiting the Machine Learning-based GAP potential.~\cite{BartokPRX2018} The simulation shown in \autoref{fig:nucleation} has been performed utilizing an integration step of 1 fs in the NPT ensemble with an initial equilibration step at zero pressure and 750 K for 20 ps. After this, a uniaxial tensile strain rate has been imposed in the vertical ($z$) direction as described in Ref.~\cite{rovaris2024unraveling}. The phase transition have been detected during the simulation exploiting a Neural Network-based phase recognition tool described in Ref.~\cite{GeActaMat2024} for the bc8 and r8 phases and a standard tool present in the OVITO software~\cite{StukowskiMSMSE2009} and based on crystalline symmetry for the dc and the hd phases. 

\textit{DFT Calculations}: First-principles calculations within the Density Functional Theory (DFT) framework were performed using the Quantum Espresso code.\cite{giannozzi2020quantum} The exchange-correlation (XC) contribution to the total energy was modeled using the Generalized Gradient Approximation (GGA) with Perdew-Burke-Ernzerhof (PBE) parametrization.\cite{perdew1996generalized} We also employed the Local Density Approximation (LDA),\cite{kohn1965self} Strongly Constrained and Appropriately Normed (SCAN)\cite{sun2015strongly, yao2017plane} assessed through the LIBXC library,\cite{lehtola2018recent} and PBE revised for solids (PBEsol)\cite{perdew2008restoring} functionals to benchmark against existing literature data. The electron-ion interaction was described using Optimized Norm-Conserving Vanderbilt (ONCV) pseudopotentials\cite{hamann2013optimized} from the PseudoDojo set.\cite{van2018pseudodojo} Wave functions were expressed using a plane-wave basis set with an energy cutoff of 100 Ry. Given the sensitivity of Raman intensities to reciprocal space sampling, a $\Gamma$-centered k-point mesh with a fine density of 0.05 Å$^{-1}$, according to the Monkhorst-Pack scheme, was used.\cite{monkhorst1976special} Structural geometry optimizations were performed via a variable cell relaxation procedure until the Hellmann-Feynman forces on each atom were less than 1 meV/Å and the stress tensor components were below 0.02 kbar. The self-consistent field iteration convergence threshold was set to 10$^{-10}$ eV. To evaluate the structural stability of the studied phases, phonon calculations were performed using density functional perturbation theory (DFPT).\cite{baroni2001phonons} First-order resonant Raman spectra for various Porto configurations were computed with the QERAMAN code.\cite{hung2024qeraman} This calculation was undertaken to elucidate specific characteristics of the measured vibrational Raman spectra and to identify the metastable phases observed experimentally. The electron energy loss spectrum (EELS) curves were obtained using the Liouville–Lanczos approach to Time-Dependent Density Functional Theory (TDDFT),\cite{timrov2013electron} implemented in the turboEELS code.\cite{timrov2015turboeels} To achieve convergence, we performed 6,000 iterations in the Lanczos recursion and extrapolated the Lanczos coefficients to 100,000 iterations using the bi-constant extrapolation method.\cite{rocca2008turbo} The EELS is obtained in the limit q$\rightarrow$0, where q is transferred momentum.

Relaxed structures were initially obtained by performing a full structural relaxation with respect to all degrees of freedom at zero pressure, utilizing various XC functionals. Subsequently, the isothermal bulk modulus (B$_0$) and its pressure derivative (B$^{'}_{0}$) were determined through a series of fixed-volume relaxations. The resulting energy-to-volume data was then fitted to the Vinet equation of state (EOS).\cite{vinet1989universal} These results, summarized in Table S2, were compared with both theoretical and experimental values available in the literature. However, it is important to acknowledge that experimental structural parameters are relatively scarce and, in some instances, derived from nanostructured or potentially strained samples. The differences in energy from the corresponding dc phase are also reported in Table S2. Our findings indicate distinct trends in lattice parameters based on functional choice. Specifically, the LDA tends to overbind, while functionals incorporating gradient corrections, particularly the PBE functional, tend to underestimate the chemical bond strength. These tendencies are also reflected in the computed volumes and bulk modulus (B$_0$). Overall, results obtained using the PBEsol and SCAN functionals exhibit better agreement with experimental data. Furthermore, our calculated structural properties are consistent with previous theoretical results, reinforcing the validity of our PBEsol and SCAN calculations as a reliable foundation for future experimental studies. On the other hand, the calculated energy differences within the SCAN functional are systematically larger than those from the PBE functional, which in turn are larger than those from the LDA and PBEsol. However, the energy differences between SCAN and the other XC functionals are quite small for the low-density tetrahedral polymorphs, with equilibrium volumes close to the dc phase, such as the hd phase. In contrast, for the high-density tetrahedral polymorphs, such as bc8 and r8 phases, there is a significant difference between the results obtained with different XC approximations. This reflects the similarity in bonding between dc and hd, and between r8 and bc8 phases. This also highlights the fact that predictions of phase stability are sensitive, in some cases strongly so, to the underlying treatment of XC effects.

The phonon band structures of the dc, hd, r8, and bc8 phases of silicon at zero pressure are plotted in Figure S9a,b,c,d. These plots demonstrate that all studied phases are dynamically stable at zero pressure, exhibiting no imaginary frequencies throughout the Brillouin zone. For the bc8 and r8 phases, the number of phonon branches is quadrupled compared to the dc phase due to the larger primitive cell, which enhances phonon scattering and reduces lattice thermal conductivity.\cite{toberer2011phonon} Additionally, the phonon spectra for the r8 phase qualitatively resemble those of the bc8 phase.

It should be noted that the consistency between our first-principles computations and experimental measurements of Raman-active mode frequencies across all studied silicon phases is due to a considered choice of the XC functional, as detailed in Table S3. After a thorough analysis of the effects of the choice of XC functional and structural parameters, we adopted an approach where PBE XC functional is used in combination with the theoretically relaxed zero pressure structures obtained using the SCAN XC functional. However, the small discrepancies between the experimental and theoretical data can be attributed to the static (0 K) calculation and the harmonic approximation employed in the Raman calculations. Furthermore, Raman spectra were computed for structures relaxed at 0 GPa, but as demonstrated in the case of dc, the observed r8, bc8, and hd phases are likely under residual pressure. This residual pressure would lead to further shifts in peak positions and changes in the Raman intensities in the computed spectra. This phenomenon was also observed by Wong et al.,\cite{wong2019formation} where bc8-Si and r8-Si phases recovered after indentation exhibited distorted unit cells, characterized by tensile strain along the indentation axis and compressive strain perpendicular to it. This strain leads to an overall reduction in unit cell volume equivalent to a quasi-hydrostatic compression of approximately 4 GPa for the same phases recovered from DAC compression.\cite{piltz1995structure} Furthermore, the crystals in the transformed zone are confirmed from TEM results to be 5–30 nm in size, implying that phonon confinement effects may shift Raman frequencies even further. Finally, Figures S10-S13 show the atomic displacements of the Raman-active phonon modes, where the arrows indicate the vibration directions of atoms.

\section*{Supporting Information}

Supporting Information is available from the Wiley Online Library or from the author.

\section*{Acknowledgements}
M.B., F.R., and E.S. acknowledge the CINECA consortium under the ISCRA initiative for the availability of high-performance computing resources and support.  
A.M.M and E.S. acknowledge financial support under the National Recovery and Resilience Plan (NRRP), Mission 4, Component 2, Investment 1.1, Call for tender No. 104 published on 2.2.2022 by the Italian Ministry of University and Research (MUR), funded by the European Union – NextGenerationEU– Project Title "SiGe Hexagonal Diamond Phase by nanoIndenTation (HD- PIT)" – CUP H53D23000780001 and B53D23004120006 - Grant Assignment Decree No. 957 adopted on 30.06.2023 by the Italian Ministry of Ministry of University and Research (MUR).

\section*{Conflict of Interest}

The authors declare no conflict of interest.


\section*{Data Availability Statement}

The data that support the ﬁndings of this study are available from the corresponding author upon reasonable request.

\section*{Keywords}
Hexagonal silicon (hd-Si), Nanoindentation, Phase transition, Textured silicon crystals, Metastable silicon phases

\bibliography{References}

\begin{thebibliography}{10}
\providecommand \doibase [0]{http://dx.doi.org/}%

\bibitem{du2021thermal}
Du PH, Zhou J. Thermal property and lattice thermal conductivity of three-dimensional pentagonal silicon. {\it Physica B: Condensed Matter.} 2021\string;618\string:413178.

\bibitem{shao2022thermoelectric}
Shao H, Ding D, Zhang L, Dong CK, Zhang H. Thermoelectric performance in a Si allotrope with ultralow thermal conductivity: a first-principles study combining phonon-limited electronic transport calculations. {\it Materials Today Physics.} 2022\string;27\string:100756.

\bibitem{liu2021thermoelectric}
Liu Z, Tan N, Tang C. Thermoelectric performance of tetragonal silicon allotrope tP 36-Si from first-principles study. {\it The European Physical Journal B.} 2021\string;94(12)\string:247.

\bibitem{zhang2018thermoelectric}
Zhang P, Ouyang T, Tang C, et al. Thermoelectric properties of four typical silicon allotropes. {\it Modelling and Simulation in Materials Science and Engineering.} 2018\string;26(8)\string:085006.

\bibitem{guo2015new}
Guo Y, Wang Q, Kawazoe Y, Jena P. A new silicon phase with direct band gap and novel optoelectronic properties. {\it Scientific reports.} 2015\string;5(1)\string:14342.

\bibitem{malone2008ab2}
Malone BD, Sau JD, Cohen ML. Ab initio study of the optical properties of Si-XII. {\it Physical Review B—Condensed Matter and Materials Physics.} 2008\string;78(16)\string:161202.

\bibitem{malone2008ab1}
Malone BD, Sau JD, Cohen ML. Ab initio survey of the electronic structure of tetrahedrally bonded phases of silicon. {\it Physical Review B—Condensed Matter and Materials Physics.} 2008\string;78(3)\string:035210.

\bibitem{he2016direct}
He C, Zhang C, Li J, et al. Direct and quasi-direct band gap silicon allotropes with remarkable stability. {\it Physical Chemistry Chemical Physics.} 2016\string;18(14)\string:9682--9686.

\bibitem{tang2015stable}
Tang CP, Cao J, Xiong SJ. Stable porous crystalline silicon with nanotubular structure: A predicted allotrope with direct band gap. {\it Physica B: Condensed Matter.} 2015\string;466\string:59--63.

\bibitem{lee2016ab}
Lee IH, Oh YJ, Kim S, Lee J, Chang KJ. Ab initio materials design using conformational space annealing and its application to searching for direct band gap silicon crystals. {\it Computer Physics Communications.} 2016\string;203\string:110--121.

\bibitem{lee2014computational}
Lee IH, Lee J, Oh YJ, Kim S, Chang KJ. Computational search for direct band gap silicon crystals. {\it Physical review B.} 2014\string;90(11)\string:115209.

\bibitem{wang2014direct}
Wang Q, Xu B, Sun J, et al. Direct band gap silicon allotropes. {\it Journal of the American Chemical Society.} 2014\string;136(28)\string:9826--9829.

\bibitem{xiang2013towards}
Xiang H, Huang B, Kan E, Wei SH, Gong X. Towards direct-gap silicon phases by the inverse band structure design approach. {\it Physical review letters.} 2013\string;110(11)\string:118702.

\bibitem{botti2012low}
Botti S, Flores-Livas JA, Amsler M, Goedecker S, Marques MA. Low-energy silicon allotropes with strong absorption in the visible for photovoltaic applications. {\it Physical Review B—Condensed Matter and Materials Physics.} 2012\string;86(12)\string:121204.

\bibitem{cohen2011wave}
Cohen ML, Malone BD. Wave function engineering: Other phases of Si for photovoltaic applications. {\it Journal of Applied Physics.} 2011\string;109(10).

\bibitem{saleev2017ab}
Saleev VA, Shipilova AV, Proserpio DM, Fadda G. Ab initio study of new sp 3 silicon and germanium allotropes predicted from the zeolite topologies. {\it The European Physical Journal B.} 2017\string;90\string:1--10.

\bibitem{he2019direct}
He Z, Maurice JL, Li Q, Pribat D. Direct evidence of 2H hexagonal Si in Si nanowires. {\it Nanoscale.} 2019\string;11(11)\string:4846--4853.

\bibitem{tang2017natural}
Tang J, Maurice JL, Fossard F, et al. Natural occurrence of the diamond hexagonal structure in silicon nanowires grown by a plasma-assisted vapour--liquid--solid method. {\it Nanoscale.} 2017\string;9(24)\string:8113--8118.

\bibitem{ahn2021optical}
Ahn HS, Kim SW, Lee GS, et al. Optical property of hexagonal (2H) silicon crystal. {\it Semiconductor Science and Technology.} 2021\string;36(9)\string:095023.

\bibitem{fadaly2020direct}
Fadaly EM, Dijkstra A, Suckert JR, et al. Direct-bandgap emission from hexagonal Ge and SiGe alloys. {\it Nature.} 2020\string;580(7802)\string:205--209.

\bibitem{peeters2024direct}
Peeters WH, Lange vVT, Belabbes A, et al. Direct bandgap quantum wells in hexagonal Silicon Germanium. {\it Nature Communications.} 2024\string;15(1)\string:5252.

\bibitem{wang2021electronic}
Wang Z, Zhang Z, Liu S, Robertson J, Guo Y. Electronic properties and tunability of the hexagonal SiGe alloys. {\it Applied Physics Letters.} 2021\string;118(17).

\bibitem{bao2021toward}
Bao N, Guo F, Kang D, Feng Y, Wang H, Dai J. Toward accurate electronic, optical, and vibrational properties of hexagonal Si, Ge, and Si1- xGex alloys from first-principle simulations. {\it Journal of Applied Physics.} 2021\string;129(14).

\bibitem{fan2020p63}
Fan Q, Zhang W, Song Y, Zhang W, Yun S. P63/mmc-Ge and their Si--Ge alloys with a mouldable direct band gap. {\it Semiconductor Science and Technology.} 2020\string;35(5)\string:055012.

\bibitem{fan2018theoretical}
Fan Q, Chai C, Wei Q, Wong K, Liu Y, Yang Y. Theoretical investigations of group IV alloys in the Lonsdaleite phase. {\it Journal of Materials Science.} 2018\string;53\string:2785--2801.

\bibitem{wong2019thermal}
Wong S, Johnson BC, Haberl B, et al. Thermal evolution of the indentation-induced phases of silicon. {\it Journal of Applied Physics.} 2019\string;126(10).

\bibitem{gupta1980static}
Gupta MC, Ruoff AL. Static compression of silicon in the [100] and in the [111] directions. {\it Journal of applied physics.} 1980\string;51(2)\string:1072--1075.

\bibitem{hu1984phases}
Hu J, Spain I. Phases of silicon at high pressure. {\it Solid state communications.} 1984\string;51(5)\string:263--266.

\bibitem{zhao2012tetragonal}
Zhao Z, Tian F, Dong X, et al. Tetragonal allotrope of group 14 elements. {\it Journal of the American Chemical Society.} 2012\string;134(30)\string:12362--12365.

\bibitem{mujica2015low}
Mujica A, Pickard CJ, Needs RJ. Low-energy tetrahedral polymorphs of carbon, silicon, and germanium. {\it Physical Review B.} 2015\string;91(21)\string:214104.

\bibitem{ge2004thermal}
Ge D, Domnich V, Gogotsi Y. Thermal stability of metastable silicon phases produced by nanoindentation. {\it Journal of Applied Physics.} 2004\string;95(5)\string:2725--2731.

\bibitem{mannepalli2019situ}
Mannepalli S, Mangalampalli KS. In-situ high temperature micro-Raman investigation of annealing behavior of high-pressure phases of Si. {\it Journal of Applied Physics.} 2019\string;125(22).

\bibitem{ruffell2007annealing}
Ruffell S, Bradby J, Williams JS. Annealing kinetics of nanoindentation-induced polycrystalline high pressure phases in crystalline silicon. {\it Applied physics letters.} 2007\string;90(13).

\bibitem{ruffell2009annealing}
Ruffell S, Haberl B, Koenig S, Bradby J, Williams JS. Annealing of nanoindentation-induced high pressure crystalline phases created in crystalline and amorphous silicon. {\it Journal of Applied Physics.} 2009\string;105(9).

\bibitem{haberl2015thermal}
Haberl B, Guthrie M, Sinogeikin SV, Shen G, Williams JS, Bradby JE. Thermal evolution of the metastable r8 and bc8 polymorphs of silicon. {\it High Pressure Research.} 2015\string;35(2)\string:99--116.

\bibitem{weill1989characterisation}
Weill G, Mansot J, Sagon G, Carlone C, Besson J. Characterisation of Si III and Si IV, metastable forms of silicon at ambient pressure. {\it Semiconductor science and technology.} 1989\string;4(4)\string:280.

\bibitem{johnson2011temperature}
Johnson B, Haberl B, Bradby J, McCallum JC, Williams J. Temperature dependence of Raman scattering from the high-pressure phases of Si induced by indentation. {\it Physical Review B—Condensed Matter and Materials Physics.} 2011\string;83(23)\string:235205.

\bibitem{smillie2020exotic}
Smillie L, Niihori M, Rapp L, et al. Exotic silicon phases synthesized through ultrashort laser-induced microexplosion: Characterization with Raman microspectroscopy. {\it Physical Review Materials.} 2020\string;4(9)\string:093803.

\bibitem{wong2019formation}
Wong S, Haberl B, Johnson B, et al. Formation of an r8-dominant Si material. {\it Physical Review Letters.} 2019\string;122(10)\string:105701.

\bibitem{hauge2015hexagonal}
Hauge HIT, Verheijen MA, Conesa-Boj S, et al. Hexagonal silicon realized. {\it Nano letters.} 2015\string;15(9)\string:5855--5860.

\bibitem{kobliska1972raman}
Kobliska R, Solin S, Selders M, et al. Raman scattering from phonons in polymorphs of Si and Ge. {\it Physical Review Letters.} 1972\string;29(11)\string:725.

\bibitem{hanfland1990raman}
Hanfland M, Syassen K. Raman modes of metastable phases of Si and Ge. {\it High pressure research.} 1990\string;3(1-6)\string:242--244.

\bibitem{olijnyk1999effect}
Olijnyk H, Jephcoat A. Effect of pressure on Raman spectra of metastable phases of Si and Ge. {\it physica status solidi (b).} 1999\string;211(1)\string:413--420.

\bibitem{kailer1999raman}
Kailer A, Nickel KG, Gogotsi YG. Raman microspectroscopy of nanocrystalline and amorphous phases in hardness indentations. {\it Journal of Raman Spectroscopy.} 1999\string;30(10)\string:939--946.

\bibitem{gerbig2015situ}
Gerbig YB, Michaels CA, Bradby JE, Haberl B, Cook RF. In situ spectroscopic study of the plastic deformation of amorphous silicon under nonhydrostatic conditions induced by indentation. {\it Physical Review B.} 2015\string;92(21)\string:214110.

\bibitem{Zarudi}
Zarudi I, Zou J, Zhang LC. {Microstructures of phases in indented silicon: A high resolution characterization}. {\it Applied Physics Letters.} 2003\string;82(6)\string:874-876.
\newblock \href {\doibase 10.1063/1.1544429} {doi: 10.1063/1.1544429}

\bibitem{crain1994reversible}
Crain J, Ackland G, Maclean J, Piltz R, Hatton P, Pawley G. Reversible pressure-induced structural transitions between metastable phases of silicon. {\it Physical Review B.} 1994\string;50(17)\string:13043.

\bibitem{piltz1995structure}
Piltz R, Maclean J, Clark S, Ackland G, Hatton P, Crain J. Structure and properties of silicon XII: A complex tetrahedrally bonded phase. {\it Physical Review B.} 1995\string;52(6)\string:4072.

\bibitem{wippermann2016novel}
Wippermann S, He Y, V{\"o}r{\"o}s M, Galli G. Novel silicon phases and nanostructures for solar energy conversion. {\it Applied Physics Reviews.} 2016\string;3(4).

\bibitem{haberl2016pathways}
Haberl B, Strobel TA, Bradby JE. Pathways to exotic metastable silicon allotropes. {\it Applied Physics Reviews.} 2016\string;3(4).

\bibitem{BartokPRX2018}
Bart{\'o}k AP, Kermode J, Bernstein N, Cs{\'a}nyi G. Machine learning a general-purpose interatomic potential for silicon. {\it Physical Review X.} 2018\string;8(4)\string:041048.

\bibitem{GeActaMat2024}
Ge G, Rovaris F, Lanzoni D, et al. Silicon phase transitions in nanoindentation: Advanced molecular dynamics simulations with machine learning phase recognition. {\it Acta Materialia.} 2024\string;263\string:119465.
\newblock \href {\doibase https://doi.org/10.1016/j.actamat.2023.119465} {doi: https://doi.org/10.1016/j.actamat.2023.119465}

\bibitem{rovaris2024unraveling}
Rovaris F, Marzegalli A, Montalenti F, Scalise E. Unraveling the Atomic-Scale Pathways Driving Pressure-Induced Phase Transitions in Silicon. {\it arXiv preprint arXiv:2408.12358.} 2024.

\bibitem{jamieson1963crystal}
Jamieson JC. Crystal structures at high pressures of metallic modifications of silicon and germanium. {\it Science.} 1963\string;139(3556)\string:762--764.

\bibitem{mcmahon1993new}
McMahon M, Nelmes R. New high-pressure phase of Si. {\it Physical Review B.} 1993\string;47(13)\string:8337.

\bibitem{mcmahon1994pressure}
McMahon M, Nelmes R, Wright N, Allan D. Pressure dependence of the Imma phase of silicon. {\it Physical Review B.} 1994\string;50(2)\string:739.

\bibitem{olijnyk1984structural}
Olijnyk H, Sikka S, Holzapfel W. Structural phase transitions in Si and Ge under pressures up to 50 GPa. {\it Physics Letters A.} 1984\string;103(3)\string:137--140.

\bibitem{butler2023origin}
Butler SL, Partridge JF, Huang X, et al. Origin of preferred orientation in an isotropic material: High pressure synthesis of bc8-Si. {\it Applied Physics Letters.} 2023\string;123(23).

\bibitem{LEITNER2018}
Leitner A, Maier-Kiener V, Kiener D. Essential refinements of spherical nanoindentation protocols for the reliable determination of mechanical flow curves. {\it Materials \& Design.} 2018\string;146\string:69-80.
\newblock \href {\doibase https://doi.org/10.1016/j.matdes.2018.03.003} {doi: https://doi.org/10.1016/j.matdes.2018.03.003}

\bibitem{Kiener2023}
Kiener D, Wurmshuber M, Alfreider M, Schaffar GJ, Maier-Kiener V. {Recent advances in nanomechanical and in situ testing techniques: Towards extreme conditions}. {\it Current Opinion in Solid State and Materials Science.} 2023\string;27(6)\string:101108.
\newblock \href {\doibase 10.1016/j.cossms.2023.101108} {doi: 10.1016/j.cossms.2023.101108}

\bibitem{lin2020temperature}
Lin C, Liu X, Yang D, et al. Temperature-and rate-dependent pathways in formation of metastable silicon phases under rapid decompression. {\it Physical Review Letters.} 2020\string;125(15)\string:155702.

\bibitem{domnich2002phase}
Domnich V, Gogotsi Y. Phase transformations in silicon under contact loading. {\it Reviews on Advanced Materials Science(Russia).} 2002\string;3(1)\string:1--36.

\bibitem{Schaffar2022}
Schaffar GJK, Kappacher J, Tscharnuter D, Maier-Kiener V. {The Phase Transformation of Silicon Assessed by an Unloading Contact Pressure Approach}. {\it JOM.} 2022\string;74(6)\string:2220--2230.
\newblock \href {\doibase 10.1007/s11837-022-05290-4} {doi: 10.1007/s11837-022-05290-4}

\bibitem{Juliano2004}
Juliano T, Domnich V, Gogotsi Y. {Examining pressure-induced phase transformations in silicon by spherical indentation and Raman spectroscopy: A statistical study}. {\it Journal of Materials Research.} 2004\string;19(10)\string:3099--3108.
\newblock \href {\doibase 10.1557/JMR.2004.0403} {doi: 10.1557/JMR.2004.0403}

\bibitem{ThompsonCPC2022}
Thompson AP, Aktulga HM, Berger R, et al. {LAMMPS} - a flexible simulation tool for particle-based materials modeling at the atomic, meso, and continuum scales. {\it Comp. Phys. Comm..} 2022\string;271\string:108171.
\newblock \href {\doibase 10.1016/j.cpc.2021.108171} {doi: 10.1016/j.cpc.2021.108171}

\bibitem{StukowskiMSMSE2009}
Stukowski A. Visualization and analysis of atomistic simulation data with OVITO--the Open Visualization Tool. {\it Modelling and Simulation in Materials Science and Engineering.} 2009\string;18(1)\string:015012.

\bibitem{giannozzi2020quantum}
Giannozzi P, Baseggio O, Bonf{\`a} P, et al. Quantum ESPRESSO toward the exascale. {\it The Journal of chemical physics.} 2020\string;152(15).

\bibitem{perdew1996generalized}
Perdew JP, Burke K, Ernzerhof M. Generalized gradient approximation made simple. {\it Physical review letters.} 1996\string;77(18)\string:3865.

\bibitem{kohn1965self}
Kohn W, Sham LJ. Self-consistent equations including exchange and correlation effects. {\it Physical review.} 1965\string;140(4A)\string:A1133.

\bibitem{sun2015strongly}
Sun J, Ruzsinszky A, Perdew JP. Strongly constrained and appropriately normed semilocal density functional. {\it Physical review letters.} 2015\string;115(3)\string:036402.

\bibitem{yao2017plane}
Yao Y, Kanai Y. Plane-wave pseudopotential implementation and performance of SCAN meta-GGA exchange-correlation functional for extended systems. {\it The Journal of chemical physics.} 2017\string;146(22).

\bibitem{lehtola2018recent}
Lehtola S, Steigemann C, Oliveira MJ, Marques MA. Recent developments in libxc—A comprehensive library of functionals for density functional theory. {\it SoftwareX.} 2018\string;7\string:1--5.

\bibitem{perdew2008restoring}
Perdew JP, Ruzsinszky A, Csonka GI, et al. Restoring the density-gradient expansion for exchange in solids and surfaces. {\it Physical review letters.} 2008\string;100(13)\string:136406.

\bibitem{hamann2013optimized}
Hamann D. Optimized norm-conserving Vanderbilt pseudopotentials. {\it Physical Review B—Condensed Matter and Materials Physics.} 2013\string;88(8)\string:085117.

\bibitem{van2018pseudodojo}
Van~Setten MJ, Giantomassi M, Bousquet E, et al. The PseudoDojo: Training and grading a 85 element optimized norm-conserving pseudopotential table. {\it Computer Physics Communications.} 2018\string;226\string:39--54.

\bibitem{monkhorst1976special}
Monkhorst HJ, Pack JD. Special points for Brillouin-zone integrations. {\it Physical review B.} 1976\string;13(12)\string:5188.

\bibitem{baroni2001phonons}
Baroni S, De~Gironcoli S, Dal~Corso A, Giannozzi P. Phonons and related crystal properties from density-functional perturbation theory. {\it Reviews of modern Physics.} 2001\string;73(2)\string:515.

\bibitem{hung2024qeraman}
Hung NT, Huang J, Tatsumi Y, Yang T, Saito R. QERaman: An open-source program for calculating resonance Raman spectra based on Quantum ESPRESSO. {\it Computer Physics Communications.} 2024\string;295\string:108967.

\bibitem{timrov2013electron}
Timrov I, Vast N, Gebauer R, Baroni S. Electron energy loss and inelastic x-ray scattering cross sections from time-dependent density-functional perturbation theory. {\it Physical Review B—Condensed Matter and Materials Physics.} 2013\string;88(6)\string:064301.

\bibitem{timrov2015turboeels}
Timrov I, Vast N, Gebauer R, Baroni S. turboEELS—A code for the simulation of the electron energy loss and inelastic X-ray scattering spectra using the Liouville--Lanczos approach to time-dependent density-functional perturbation theory. {\it Computer Physics Communications.} 2015\string;196\string:460--469.

\bibitem{rocca2008turbo}
Rocca D, Gebauer R, Saad Y, Baroni S. Turbo charging time-dependent density-functional theory with Lanczos chains. {\it The Journal of Chemical Physics.} 2008\string;128(15).

\bibitem{vinet1989universal}
Vinet P, Rose JH, Ferrante J, Smith JR. Universal features of the equation of state of solids. {\it Journal of Physics: Condensed Matter.} 1989\string;1(11)\string:1941.

\bibitem{toberer2011phonon}
Toberer ES, Zevalkink A, Snyder GJ. Phonon engineering through crystal chemistry. {\it Journal of Materials Chemistry.} 2011\string;21(40)\string:15843--15852.

\end{thebibliography}

\end{document}